\documentclass[11pt]{article}

\usepackage{geometry}
\usepackage[square,numbers]{natbib}
\usepackage{amsfonts,amsmath,amssymb,amsthm}
\usepackage{graphicx}
\usepackage[colorlinks,citecolor=green,linkcolor=blue,bookmarks=false,hypertexnames=true]{hyperref}
\usepackage{mathrsfs}

\allowdisplaybreaks

\def\const{\mathrm{const}}
\def\pct{\%}

\begin{document}

\title{On a priori bounding the growth of thermal instability waves}

\author{F.J.\ Beron-Vera\\ Department of Atmospheric Sciences\\ Rosenstiel School of Marine, Atmospheric \& Earth Science\\ University of Miami\\ Miami, Florida, USA\\ fberon@miami.edu}
\date{Started: January 26, 2024. This version: \today.\vspace{-0.25in}}
\maketitle

\begin{abstract}
We have previously shown that the nonlinear growth of a finite-amplitude perturbation to a basic state given by a baroclinic zonal flow on the $\beta$-plane in a thermal quasigeostrophic reduced-gravity model can be a priori bounded.  In this note we show that, unlike we stated earlier, Lyapunov stability can be proved even when buoyancy varies linearly with the meridional coordinate.  In addition to rectifying our earlier results we expand them by deriving an instability saturation bound by making use of the existence of such a class of Lyapunov-stable basic states.  This bound can be smaller than that one we estimated before, reinforcing our previous conclusions.  We also present a numerical test of the accuracy of the derived bound.
\end{abstract}

In this note we review and expand our previous results \cite{Beron-21-POFa} on the nonlinear saturation of so-called thermal instabilities, namely, those that two-dimensional quasigeostrophic flow can undergo when buoyancy (temperature) is allowed to vary with horizontal position and time.  Such thermal instabilities are the subject of active research for offering a simplified description of small-scale Kelvin--Helmholtz-like vortex rolls often seen in satellite images of the ocean surface.  Our earlier paper \cite{Beron-21-POFa} includes all pertinent references.  We include only the most relevant ones in this note, which has been made sufficiently selfconsistent to facilitate its reading.  

We begin by recalling the thermal quasigeostrophic model in a reduced-gravity or equivalent-barotropic setting, i.e., with a layer of active fluid floating atop an infinitely deep inert one \cite{Ripa-RMF-96}, a member of a more general class of thermal quasigeostrophic models recently proposed \cite{Beron-21-POFb}. In the notation of \cite{Beron-21-POFa}, which we follow closely, the equations of the model are:
\begin{subequations}
\begin{equation}
    \partial_t\bar\xi + {\{\bar\psi,\bar\xi - R^{-2}\psi_\sigma\}} = 0,\quad
    \partial_t\psi_\sigma + {\{\bar\psi,\psi_\sigma\}} = 0,
\end{equation}
with $\bar\psi(\mathbf x,t)$ depending nonlocally on $\bar\xi,\psi_\sigma(\mathbf x,t)$ through
\begin{equation}
    \nabla^2\bar\psi - R^{-2} \bar\psi = \bar\xi - R^{-2}\psi_\sigma - \beta y.
    \label{eq:inv}
\end{equation}
\label{eq:IL0QG}%
\end{subequations}
Here ${\{a,b\}} := \nabla^\perp a\cdot \nabla b$ for any $a,b(\mathbf x,t)$ where $\mathbf x = (x,y)$ ranges in a periodic zonal channel $\mathscr D$ of the $\beta$-plane of length (period) $L$ and width $W$.  Boundary conditions are $\partial_x\bar\psi\vert_{y=0,W} = 0$ (no flow normal to the channel coasts) and $\smash{\int_0^L}\partial_y\bar\psi\vert_{y=0,W}\,dx = \const$ (constancy of Kelvin circulations along channel coasts). Parameter $R := \sqrt{g'H}/|f_0|$ is the Rossby radius of deformation of the system, where $H$ and $g'$ are reference depth and buoyancy, respectively, and $f_0$ is the mean Coriolis parameter.  In a model with arbitrary stratification, $R$ corresponds, essentially, to the first internal deformation radius. 

The instantaneous layer thickness is $H (1 + (\bar\psi - \psi_\sigma)/f_0R^2)$ and the buoyancy $g' (1 + 2\psi_\sigma/f_0R^2)$, which is materially conserved, i.e., it is advected (by Lie transport as a scalar) under the flow of $\nabla^\perp\bar\psi$, the (horizontal) velocity of the thermal quasigeostrophic model. While the velocity formally represents a (nonautonmous) two-dimensional vector field, it includes implicitly by the thermal-wind balance a vertical shear, given by $2\nabla^\perp\psi_\sigma/H$. In turn, the quantity $\bar\xi$ represents the potential vorticity of the model, which, unlike that one in adiabatic (i.e., with constant density) quasigeostrophic theory, does not constitute a materially conserved quantity. Modulo some factors and with an error of the order of the Rossby number squared, $\bar\xi$ corresponds to the \emph{vertically averaged} Ertel $z/h$-potential vorticity of the primitive equations defined in $0 \le z \le -h(\mathbf x,t)$, which is \emph{not} materially conserved \cite{Ripa-JFM-95}.  Production of $\bar\xi$ under advection leads to production of Kelvin circulations along material loops, which has been argued \cite{Holm-etal-21} to be responsible for a tendency of the model to develop submesoscale, i.e., with radii smaller than $R$, vortices rolling up along fronts.

System \eqref{eq:IL0QG} has several integrals of motion, namely, energy, zonal momentum, and an infinite family of Casimirs, respectively given by
\begin{equation}
\mathcal E[\bar\xi,\psi_\sigma] := -\frac{1}{2} \int \bar\psi\bar\xi,\quad
\mathcal M[\bar\xi,\psi_\sigma] := \int y\bar\xi,\quad
\mathcal C_{C_1,C_2}[\bar\xi,\psi_\sigma] := \int C_1(\psi_\sigma) + \bar\xi C_2(\psi_\sigma)
\end{equation}
modulo Kelvin circulations along the coasts with $C_1,C_2$ being arbitrary functions, where $\int$ is short for integral over $\mathscr D$.  Let $\bar U$ and $U_\sigma$ be two constants with units of velocity. Consider the general conservation law
\begin{equation}
    \mathcal H_{\bar U} := \mathcal E - \bar U\mathcal M + \mathcal C_{0,-\tau\psi_\sigma^2/2R^2}.
\end{equation}
Consider further 
\begin{equation}
    \bar\Xi = \big(R^{-2}(\bar U - U_\sigma) + \beta\big)y,\quad
    \Psi_\sigma = - U_\sigma y,\quad
    \label{eq:BS}
\end{equation}
namely, a basic state (steady solution or equilibrium) of \eqref{eq:IL0QG} representing (implicitly) a baroclinic zonal current on the $\beta$-plane with meridionally \emph{linear} buoyancy (model variables are capitalized when representing basic state variables throughout the paper).  The condition $U_\sigma < f_0R^2/2W$ guarantees static stability.  As in \cite{Beron-21-POFa}, we compute
\begin{equation}
    \Delta \mathcal H_{\bar U}[\delta\bar\xi,\delta\psi_\sigma](\tau) =  \frac{1}{2} \int |\nabla\delta\bar\psi|^2  + R^{-2} \left(\delta\bar\psi^2 - \tau \delta\psi_\sigma^2\right),
    \label{eq:P}
\end{equation}
where $(\delta\bar\xi,\delta\psi_\sigma)$ is a perturbation to \eqref{eq:BS} of finite amplitude and arbitrary structure, and
\begin{equation}
    \tau := \bar U/U_\sigma.
\end{equation}
Sometimes referred to as a pseudo-energy--momentum, \eqref{eq:P} is preserved under the fully nonlinear dynamics produced by \eqref{eq:IL0QG}.

Positive definiteness of \eqref{eq:P}, guaranteed under the condition
\begin{equation}
   \tau < 0,
   \label{eq:stable}
\end{equation}
implies Lyapunov stability of \eqref{eq:BS}, unlike we stated in \cite{Beron-21-POFa}.  Specifically, let $\eta$ and $\lambda$ be two positive constants respectively such that
\begin{equation}
   0 < \eta < -\tau
   \label{eq:eta}
\end{equation}
and
\begin{equation}
   -\tau - \eta \le \lambda \le -\tau + \eta.
   \label{eq:lambda}
\end{equation}
Then the convexity estimate readily follows:
\begin{equation}
   \frac{-\tau-\eta}{\lambda} \|(\delta\bar\xi,\delta\psi_\sigma)\|_2(\lambda)^2 \le \Delta \mathcal H_{\bar U}[\delta\bar\xi,\delta\psi_\sigma](\tau) \le \frac{-\tau+\eta}{\lambda} \|(\delta\bar\xi,\delta\psi_\sigma)\|_2(\lambda)^2,
   \label{eq:conv}
\end{equation}
where  
\begin{equation}
    \|(\delta\bar\xi,\delta\psi_\sigma)\|_2(\lambda) := \sqrt{\Delta \mathcal H_{\bar U}[\delta\bar\xi,\delta\psi_\sigma](-\lambda)}
    \label{eq:L2}
\end{equation}
measures the distance to \eqref{eq:BS}, in an $L^2$ sense in the infinite-dimensional phase space of system \eqref{eq:IL0QG} with coordinates $(\bar\xi,\psi_\sigma)$. This establishes \cite{Holm-etal-85} Lyapunov stability for \eqref{eq:BS}.  Indeed, since \eqref{eq:P} is invariant,  from \eqref{eq:conv} it follows that the distance to \eqref{eq:BS} at any time is bounded from above by a multiple of the initial distance. {More explicitly, upon evaluating the left (resp., right) inequality in \eqref{eq:conv} at $t>0$ (resp., $t=0$), one gets
\begin{equation}
    \left.\|(\delta\bar\xi,\delta\psi_\sigma)\|_2(\lambda)\right\vert_{t>0} \le \sqrt{\frac{\tau-\eta}{\tau+\eta}} \left.\|(\delta\bar\xi,\delta\psi_\sigma)\|_2(\lambda)\right\vert_{t=0}.
    \label{eq:lyapunov}
\end{equation}
This implies Lyapunov stability since for every $\varepsilon > 0$, there exists $\mu(\varepsilon) >0$ such that $\left.\|(\delta\bar\xi,\delta\psi_\sigma)\|_2(\lambda)\right\vert_{t=0} < \mu$ implies $\left.\|(\delta\bar\xi,\delta\psi_\sigma)\|_2(\lambda)\right\vert_{t>0} < \varepsilon$, which is satisfied upon choosing $\mu = \sqrt{(\tau + \eta)/(\tau - \eta)}\varepsilon$.}

Let us now turn to seek to a priori {constraining} the growth of thermal instability waves \cite{Gouzien-etal-17}, i.e., normal-mode perturbations to basic states \eqref{eq:BS} that necessarily violate \eqref{eq:stable}, by making use of the above stability result.  Indeed, such normal modes can grow (and decay) when their wavelengths are short enough; in particular, when $\tau = 1$, this happens independent of the wavenumber \cite{Beron-21-POFa}. Denote stable (resp., unstable) state quantities with a superscript $\mathrm{S}$ (resp., $\mathrm{U}$).  Grouping $(\bar\xi,\psi_\sigma)$ into $\varphi$ we have, as discussed in \cite{Beron-21-POFa},
\begin{equation}
   \|\varphi - \Phi^\mathrm{U}\|_2(\lambda) \le \sqrt{\frac{2\tau^\mathrm{S}}{\tau^\mathrm{S}+\eta}} \|\Phi^\mathrm{S} - \Phi^\mathrm{U}\|_2(\lambda)
   \label{eq:shepherd}
\end{equation}
as a result of successively \cite{Shepherd-88a} 1) applying the triangular inequality $\|\varphi - \Phi^\mathrm{U}\|_2(\lambda) \le \|\varphi - \Phi^\mathrm{S}\|_2(\lambda) + \|\Phi^\mathrm{S} - \Phi^\mathrm{U}\|_2(\lambda)$, 2) using {Lyapunov stability statement \eqref{eq:lyapunov}}, and 3) \emph{assuming that $\varphi \approx \Phi^\mathrm{U}$ initially at $t = 0$}.  Upon evaluating the right-hand-side of \eqref{eq:shepherd}, we obtain the following upper bound on the nonlinear growth of a finite-amplitude perturbation to a basic state \eqref{eq:BS} violating \eqref{eq:stable}, i.e., possibly unstable, as measured by \eqref{eq:L2}:
\begin{equation}
    \mathcal B := \sqrt{\frac{2\bar U^\mathrm{S}}{\bar U^\mathrm{S}+\eta U_\sigma^\mathrm{S}}} \cdot \sqrt{\frac{LW^3}{6R^2}} \cdot \sqrt{(\bar U^\mathrm{S} - \bar U^\mathrm{U})^2 - \left(\frac{\bar U^\mathrm{S}}{U_\sigma^\mathrm{S}} + \eta\right) (U_\sigma^\mathrm{S} - U_\sigma^\mathrm{U})^2},
\end{equation}
where we have set $\lambda$ to its minimum value in the admissible range \eqref{eq:lambda}. 

Let us take
\begin{equation}
    \eta = -r\bar U^\mathrm{S}/U_\sigma^\mathrm{S},\quad 0 < r < 1,
\end{equation}
satisfying \eqref{eq:eta}.
Then
\begin{equation}
    \mathcal B = \sqrt{\frac{2}{1-r}} \cdot \sqrt{\frac{LW^3}{6R^2}} \cdot \sqrt{(\bar U^\mathrm{S} - \bar U^\mathrm{U})^2 + (r-1) \frac{\bar U^\mathrm{S}}{U_\sigma^\mathrm{S}} (U_\sigma^\mathrm{S} - U_\sigma^\mathrm{U})^2}.
    \label{eq:B}
\end{equation}
A bound tighter than \eqref{eq:B} follows by replacing the leftmost square root in it with its minimum value as a function of $r$, given by $\sqrt{2}$, and doing likewise with $r - 1$ in the rightmost square root, namely, $-1$. The result is
\begin{equation}
    \mathcal B \ge \hat{\mathcal B} := \sqrt{\frac{LW^3}{3R^2}} \cdot \sqrt{(\bar U^\mathrm{S} - \bar U^\mathrm{U})^2 - \frac{\bar U^\mathrm{S}}{U_\sigma^\mathrm{S}} (U_\sigma^\mathrm{S} - U_\sigma^\mathrm{U})^2}.
    \label{eq:Bhat}
\end{equation}
Using Matlab's Symbolic Math Toolbox, we find
\begin{equation}
    \hat{\mathcal B} \ge \mathcal B_\mathrm{opt} :=  \min_{(\bar U^\mathrm{S},U_\sigma^\mathrm{S}) \neq (\bar U^\mathrm{U},U_\sigma^\mathrm{U})}\hat{\mathcal B} = 2\sqrt{\frac{LW^3}{3R^2}}\cdot |U_\sigma^\mathrm{U}|\sqrt{\frac{\bar U^\mathrm{U}}{U_\sigma^\mathrm{U}} - 1},
    \label{eq:Bopt}
\end{equation}
being attained at $(\bar U^\mathrm{S}, U_\sigma^\mathrm{S}) = (\bar U^\mathrm{U}-2U_\sigma^\mathrm{U}, -U_\sigma^\mathrm{U})$ conditioned on $\bar U^\mathrm{U}/U_\sigma^\mathrm{U} > 2$ (a tractable closed-form formula valid globally seems impossible to be calculated).

\begin{figure}[p!]
    \centering
    \includegraphics[width=.75\textwidth]{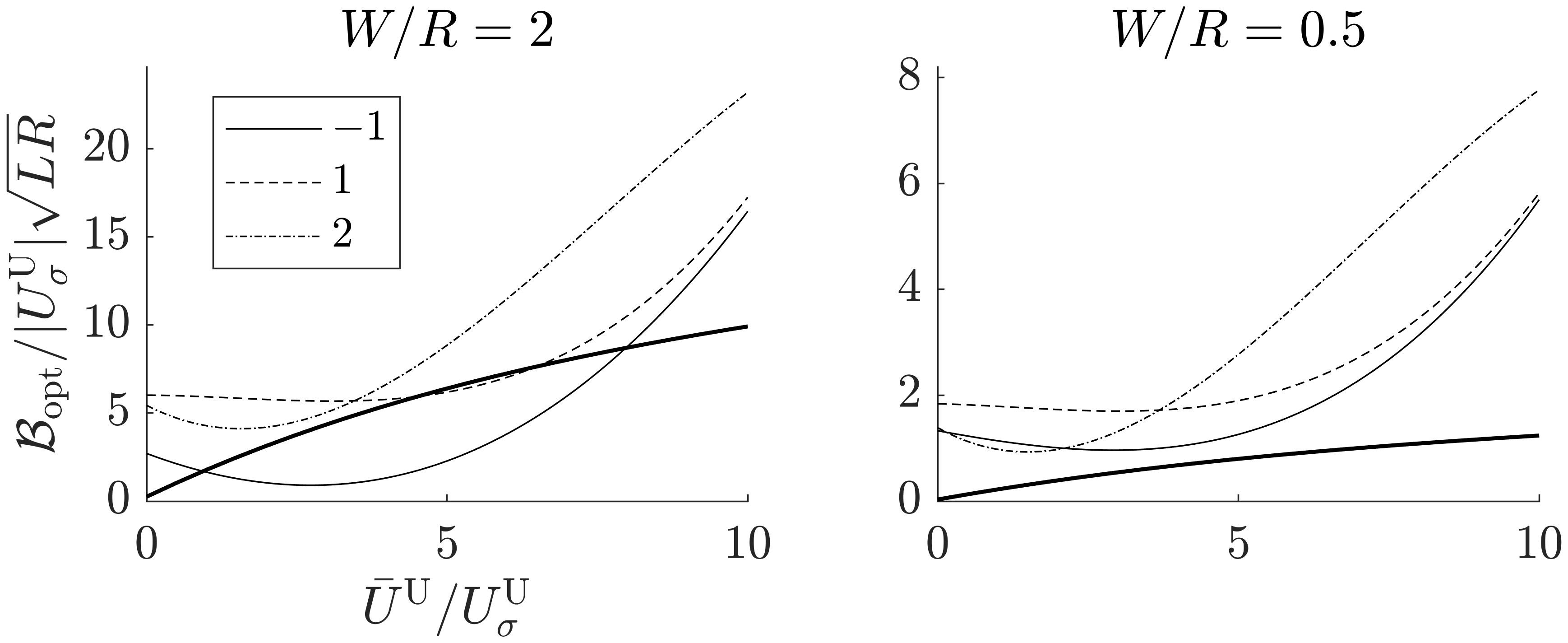}
    \caption{Nonlinear saturation bound on the growth of perturbations to a basic state \eqref{eq:BS} violating \eqref{eq:stable}, as obtained here (heavy) and as computed in \cite{Beron-21-POFa} (light, various line styles) for selected $W/R$ values. The basic state represents, implicitly, a baroclinic zonal current on the $\beta$-plane with meridionally linear buoyancy (cf.\ text for details). Each number in the legend corresponds to a selected value of $U_\sigma^\mathrm{U}$ as measured by $f_0R$.}
    \label{fig:bound}
\end{figure}

In Fig.\ \ref{fig:bound} we show as a heavy line $\mathcal B_\mathrm{opt}$, as expressed analytically in \eqref{eq:Bopt} for $\bar U^\mathrm{U}/U_\sigma^\mathrm{U} > 2$ or as estimated numerically otherwise, as a function of $\bar U^\mathrm{U}/U_\sigma^\mathrm{U}$ for the two $W/R$ values considered in \cite{Beron-21-POFa}.  The corresponding results based on the bound derived in \cite{Beron-21-POFa} using a Lyapunov-stable basic state representing a baroclinic zonal jet with a nonlinear buoyancy distribution, Eqs.\ (10) and (14) in \cite{Beron-21-POFa}, are depicted as light lines of several styles.  These are the results for selected $U_\sigma^\mathrm{U}$ as measured by $f_0R$, on which the normalized bound $\hat{\mathcal B}/|U_\sigma^\mathrm{U}|\sqrt{LR}$, as evaluated using the one derived in \cite{Beron-21-POFa}, depends.  The bound computed here vanishes at criticality, $\bar U^\mathrm{U}/U_\sigma^\mathrm{U} = 0$, as can be desired.  Moreover, this can be tighter than that one computed in \cite{Beron-21-POFa}, thereby both improving its results and reinforcing its conclusions.  That is, the production of wave activity in the thermal quasigeotrophic model is constrained to the extent that it cannot undergo an ultraviolet explosion, this despite spectral theory predicts no high-wavenumber instability cutoff \cite{Beron-21-POFa}.

\begin{figure}[p!]
    \centering
    \includegraphics[width=.75\textwidth]{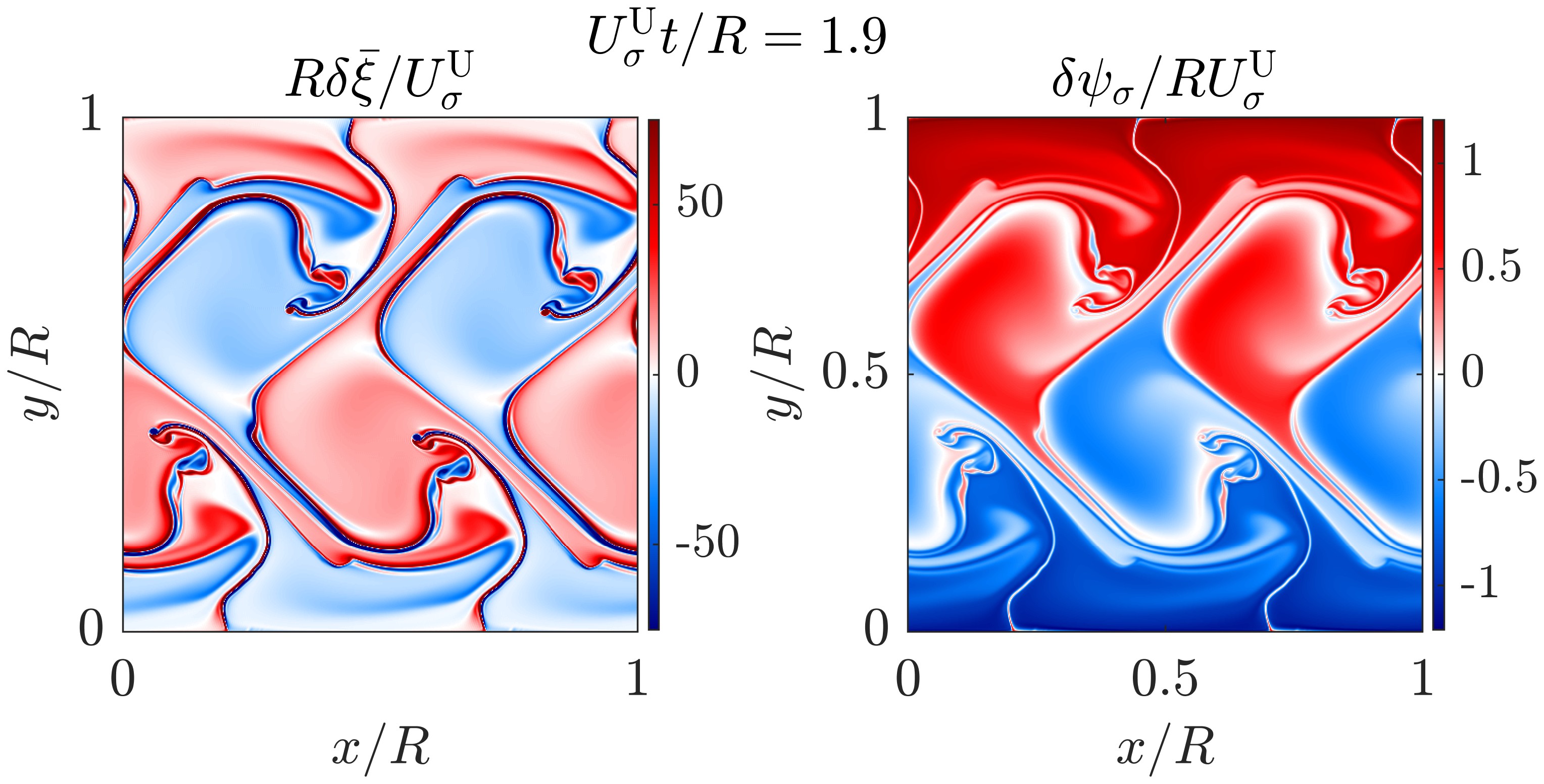}\\
    \includegraphics[width=.75\textwidth]{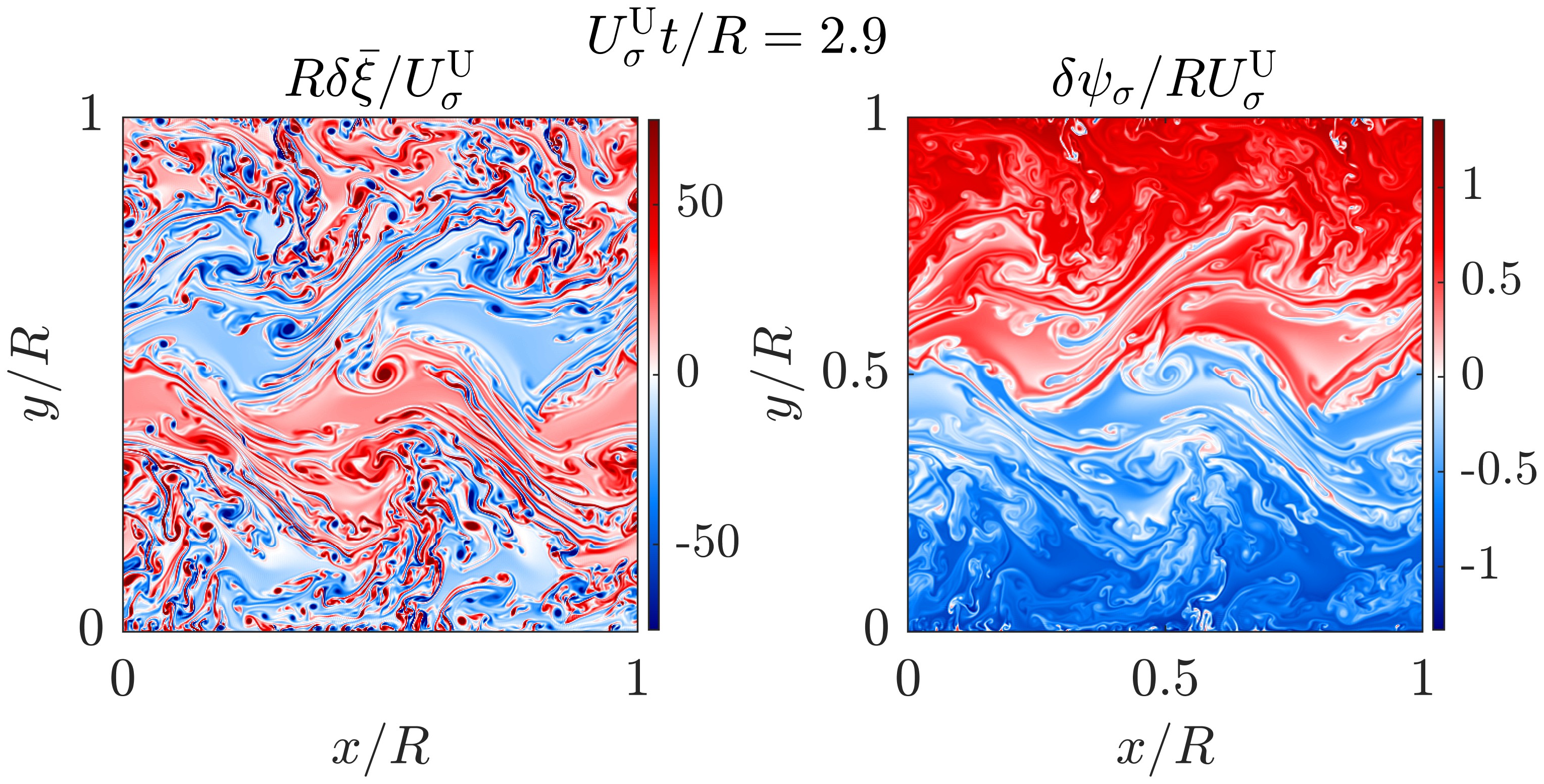}\\\vspace{.15cm}
    \includegraphics[width=.75\textwidth]{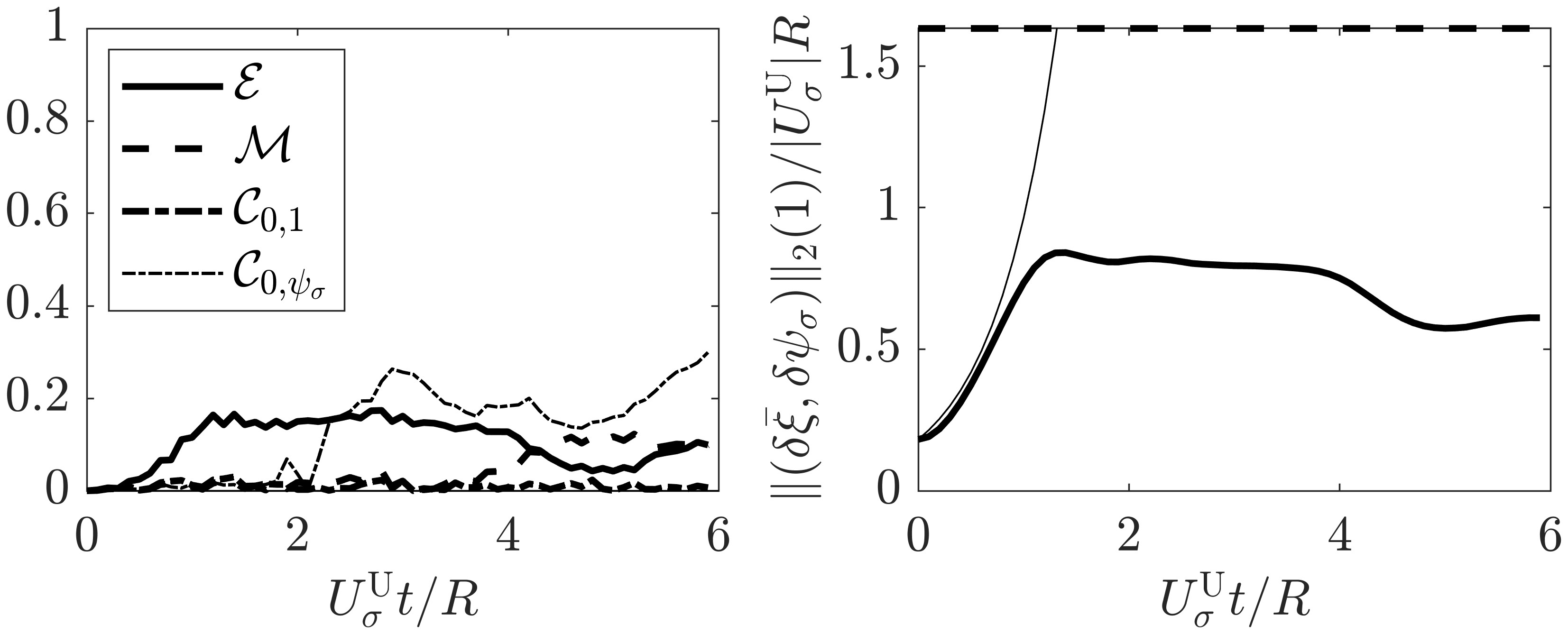}
    \caption{(top-two rows) Selected snapshots of the evolution of perturbation potential vorticity (left) and buoyancy (right) to a basic state \eqref{eq:BS} violating \eqref{eq:stable}.  (bottom-left panel) Absolute relative error with respect to the initial value of the energy, zonal momentum, and two selected Casimirs of the infinite family. (bottom-right panel) Evolution of the $L^2$ distance \eqref{eq:L2} to the unstable basic state (thick solid).  The dashed line represents an a priori upper bound on thermal-instability-wave growth as deduced using the existence of Lyapunov-stable basic states.  The thin line represents the growth as determined by spectral stability theory.}
    \label{fig:simulation}
\end{figure}

We close this note with an admittedly very modest numerical exercise aimed at testing the accuracy of the derived a priori bound on nonlinear thermal-instability-wave growth (numerical details of this exercise are deferred to the last paragraph of this note).  To the best of our knowledge, this has so far never been attempted.  The test is done by carrying out a direct numerical simulation of the IL$^0$QG set \eqref{eq:IL0QG}, initialized closed to an unstable basic state \eqref{eq:BS}.  We specifically have chosen to consider $\tau^\mathrm{U} = \bar U^\mathrm{U}/U_\sigma^\mathrm{U} = 3$, which violates Lyapunov's sufficient stability condition \eqref{eq:stable}.  This falls within the range of validity of optimal bound's explicit formula \eqref{eq:Bopt}. Taking the zonal channel period ($L$) and width ($W$) to be equal to the deformation radius ($R$), as we have chosen to do in our simulation, we get $\mathcal B_\mathrm{opt}/|U_\sigma^\mathrm{U}|R = \sqrt{8/3}$.  Moreover, the growth of the perturbation can then be consistently measured using \eqref{eq:L2} with $\lambda = -\tau^\mathrm{S} = \tau^\mathrm{U} - 2 = 1$, which leads to \eqref{eq:Bopt} upon minimization of \eqref{eq:Bhat}.  The initial perturbation is taken to be a small-amplitude normal mode.  Since Lyapunov stability and the a priori instability bound do not depend on the Charney number ($b:=\beta U_\sigma/R^2$), we have set it to be equal to zero in our simulation.  Specifically, 
\begin{equation}
    \frac{R\delta\bar\xi}{U_\sigma^\mathrm{U}} =  -\frac{\delta\psi_\sigma}{RU_\sigma^\mathrm{U}} = \frac{1}{2}\cos 4\pi \frac{x}{R}\sin \pi \frac{y}{R}.
\end{equation}
The amplitude of the perturbations respectively represent half and minus half the maximum (in absolute magnitude as it applies) unstable basic state values in the channel domain, namely, $\max_{y\in[0,W]}\bar\Xi^\mathrm{U}(y)R/U_\sigma^\mathrm{U} = (\tau^\mathrm{U} + b^\mathrm{U} - 1)W/R = 2$ and $\max_{y\in[0,W]}|\Psi_\sigma^\mathrm{U}(y)|/RU_\sigma^\mathrm{U} = W/R = 1$. Spectral (i.e., linear) stability (cf.\ Eq.\ (6) of \cite{Beron-21-POFa}) predicts a nondimensional growth rate, $\omega U_\sigma^\mathrm{U}/R$, for wavenumber-(2,1) $\mathbf k = (k,l) = (4\pi R^{-1},\pi R^{-1})$ and the above basic state parameters of $kR \operatorname{Im}\sqrt{(1 + \tau^\mathrm{U} + b)^2 - 4\tau^\mathrm{U}(|\mathbf k|^2R^2 + 1)}/(2|\mathbf k|^2R^2 + 2)) \approx 1.7.$  The top-two rows of Fig.\ \ref{fig:simulation} show two snapshots of $\delta\bar\xi$ (perturbation potential vorticity) and $\delta\psi_\sigma$ (perturbation buoyancy) at intermediate times throughout a $10\omega^{-1}$-long simulation ($\omega^{-1}$ represents one e-folding time as determined by spectral stability theory).  Note the development of small-scale Kelvin--Helmholtz-like vortex rolls.  These are similar to those reported in earlier simulations, but as a result of the imposition of \emph{topographic forcing} \cite{Beron-21-POFa,Beron-21-POFb,Holm-etal-21,Crisan-etal-23}.  In the bottom-right panel of Fig.\ \ref{fig:simulation} we plot, in solid thick, the $L^2$ distance \eqref{eq:L2} to the unstable basic state as a function of time.  The thin curve represents the growth as predicted by spectral stability theory, which is followed quite closely by the fully nonlinear evolution during its early stages.  The spectral (linear) result predicts unbounded growth, which is actually halted during the nonlinear evolution.  The growth of the thermal-instability-wave amplitude reaches a maximum at $t \approx U_\sigma/R$, which represents about 52\pct\, of the predicted \emph{possible} maximum growth, indicated by the dashed line in the bottom-right panel of Fig.\ \ref{fig:simulation}.  The theoretical prediction can be fairly considered to be quite good given the modesty of our simulation.  The confidence on its results is measured by how well the conservation laws of the system are numerically represented (Fig.\ \ref{fig:simulation}, bottom-left panel).  The relative error with respect to the initial value of the energy and zonal momentum are kept, in absolute magnitude, within less than roughly 17\pct\, and 12\pct, respectively.  The $C_{0,1}$ Casimir, namely, the integral of $\bar\xi$, is much better preserved.  The absolute relative error is maintained below about 3\pct\, along the simulation.  The family of invariant Casimirs is infinite, thus its conservation can never be directly verified entirely. An important Casimir is $C_{0,\psi_\sigma}$ as this is used to construct the pseudo-energy--momentum integral \eqref{eq:P}, involved in the calculation of Lyapunov stability and the derivation of the a priori nonlinear saturation bound on instability.  This Casimir is not conserved.  The absolute relative error increases with time; by the end of the simulation it had increased by nearly 30\pct.  More precisely, the sign of the relative error is negative, meaning that $C_{0,\psi_\sigma}$ decayed with time.   A more detailed test of the accuracy of the theoretical upper bound on allowable wave growth is deemed needed, exhaustively covering the space of parameters.  This will ideally require one to make use of a geometry-preserving numerical method. At the time of writing, this still awaits to be derived for flows on the annulus (periodic channel), domain of interest here.  A final comment is that the a priori bound derived here does not seem possible to be improved further.  In adiabatic QG flow theory a possibility exists because the convexity estimate on the pseudo-energy--momentum integral can be made tighter, leading to a tighter bound \cite{Shepherd-88b, Olascoaga-Ripa-99}.  Such flows are special inasmuch as the $L^2$ distance is given by the perturbation enstrophy, which is bounded from above by the total enstrophy of the system, conserved under the dynamics.  The IL$^0$QG does not preserve enstrophy, yet it produces phenomena (small-scale circulations) akin to those observed in nature, which the adiabatic theory cannot produce.

The numerical simulation employed a pseudosepectral scheme with fast Fourier transform (FFT) in $x$ and discrete sine transform (DST) in $y$ to invert \eqref{eq:inv}, but as written for the \emph{perturbation} to the (unstable) basic state. We consistently wrote the entire set \eqref{eq:IL0QG} accordingly. The DST imposes a homogeneous Dirichlet boundary condition on $\delta\bar\psi$ without spoiling the strict zero-normal-flow at the channel northern and southern coasts boundary conditions, namely, $\partial_x\bar\psi\vert_{y=0,W} = 0$.  A total of 512 grid points in each direction, zonal and meridional, was considered.  Differentiation was done using dialiased FFT in each direction using the $\frac{3}{2}$ zero-padding rule.  The equations were forward advanced using a fourth-order Runge--Kutta method with time step $\Delta t U_\sigma^\mathrm{U}/R \approx 0.0001$ as resulting by applying the Courant--Friedrichs--Lewy condition.  Finally, a small amount (roughly $-1.5\times 10^{-11}U_\sigma^\mathrm{U}R^3$) of biharmonic hyperviscosity was included to stabilize the time step.  

\section*{Acknowledgments}

Gage Bonner independently verified Eq.\ \eqref{eq:Bopt}.

\section*{Author declarations}

\subsection*{Conflict of interest}

The author has no conflicts to disclose.

\subsection*{Author contributions}

This paper is authored by a single individual who entirely carried out the work.

\section*{Data availability}

Data sharing is not applicable to this article as no new data were created or analyzed in this study.

\bibliographystyle{plain}

\end{document}